\documentstyle[12pt]{article}
\begin{document}

\begin{titlepage}
{\hspace*{\fill} \small Alberta-Thy-19-93 \\
\hspace*{\fill} \small hep-th 9305006}

\begin{center}
{\large\bf Two dimensional general covariance \\
\large\bf from three dimensions}
\end{center}

\begin{center}
{\small Viqar Husain\\
\small Theoretical Physics Institute, University of Alberta, \\
\small Edmonton, Canada T6G 2J1}
\end{center}
\medskip
\centerline{\small March 1993}
\medskip
\begin{abstract}
A 3d generally covariant field theory having some unusual
properties is described. The theory has a degenerate
3-metric which effectively makes it a 2d field theory in
disguise. For 2-manifolds without boundary, it has an infinite
number of conserved charges that are associated with graphs in two
dimensions and the Poisson algebra of the charges is closed.
For 2-manifolds with  boundary there are additional observables
that have a Kac-Moody Poisson algebra. It is further shown that the
theory is classically integrable and the general solution of the
equations of motion is given.
  The quantum theory is described using Dirac quantization, and it is
  shown that there are quantum states associated with graphs in two
dimensions.
\end{abstract}
\end{titlepage}
\vfill
\eject

The interest in quantum gravity is often focused on various
generally covariant toy models. These range from lower dimensional
 toy strings and matrix models \cite{gins} to
lower dimensional \cite{3dgr} and  mini/midi-superspace reductions of
general relativity \cite{minsup}, as well as assorted couplings of
these
to matter fields.

 One of the questions asked is what the
gauge invariant observables are, since these are the natural
objects to represent as Hermitian operators on the physical
state space. In a generally covariant theory, this question is
inextricably linked to the constants of the motion of the theory
and hence to integrability. This is because evolution in time
in the Hamiltonian picture is generated by the Hamiltonian
constraint which is also the constraint associated with the time
reparametrization invariance of the theory. Thus {\it fully} gauge
invariant observables are also constants of the motion.

In this regard it is of interest to ask if there are
generally covariant field theories with true local degrees
of freedom that are integrable. Theories which have only a finite
number of degrees of freedom, such as topological theories \cite{hor}
and some Bianchi models \cite{ashtu}, provide simpler examples where
fully gauge invariant observables can be extracted.

In this paper an example of a 3d generally covariant
field theory is given which is integrable in the sense that the
 explicit  general solution of the equations of motion can be
 written down. This arises mainly because the theory is
 effectively two dimensional.

 The action was found by asking whether there are any generally
 covariant (non-topological) field theories in which there is no
 dynamics. From a Hamiltonian point of view, this means that the
 initial data for the theory on a Cauchy surface, subject possibly to
 some first class constraints reflecting other symmetries, does not
evolve
in `time'. Since evolution in generally covariant theories is
generated
 by a Hamiltonian constraint, no evolution means the absence of this
 constraint in the Hamiltonian theory. The initial
 conditions that are solutions of any other `kinematical' constraints
 would then be the solutions of the `equations of motion'.
 It is in this sense that the theory would be of one lower dimension.

Such a theory is known in four
dimensions \cite{kv}, and its action is like that for general
relativity except that the internal gauge group is SO(3) instead of
 SO(3,1). This choice of gauge group makes the 4-metric degenerate
 with signature (0,+,+,+). It is this property that is manifested in
 the Hamiltonian theory by the missing Hamiltonian constraint.
The phase space variables thus do not change
 from one `spatial' surface to the next (apart from the kinematical
 gauge transformations), and so are independent of `time'.
 Thus the theory is effectively 3-dimensional.

This 4d model was in turn motivated by the program
for the canonical quantization of gravity  using the
Ashtekar variables \cite{ash}, and it captures the kinematical
features of that formalism. Similar models with
matter couplings \cite{carlo,lee,vh} have since been discussed for
the purpose of studying diffeomorphism invariant observables
in generally covariant field theories.

Of related  interest is the work relating  2d physics
to 3d topological theories. The basic idea here involves
looking at a topological theory on a 3-manifold with
boundary and asking  what 2-dimensional field theory is induced
on this boundary \cite{witten,carlip}. It is known that the
3-dimensional
Chern-Simons theories induce WZNW models on the boundary, and this
has been studied from both the path-integral and  Hamiltonian points
of view \cite{carlip,bal}.  It has been suggested
that the 2-boundary may be viewed as a string world sheet, and that
pushing the analogy may allow the calculation of string amplitudes
directly from the 3-dimensional topological theory. (For a review see
ref. \cite{carlip}).
 The WZNW theories arise from the topological Chern-Simons theory on
3-manifolds with boundary by first noting that the variation of the
 CS action gives a boundary contribution. For the variational
principle
 to be well defined a surface term must be added to the action
(analogous to the situation for general relativity). This  modified
 action is however not gauge invariant: gauge transformations
generate
 an additional surface term. This surface term is the action
 for a WZNW model.

 In the following a 3d generally
covariant field theory on a manifold $M$ is described which is
effectively a 2d field theory in the sense described above, namely
 that it has no Hamiltonian constraint.
The Hamiltonian version of  the theory on $M=\Sigma\times R$
is discussed in some detail and it is shown that there are an
infinite number
of conserved charges for both compact and noncompact 2-manifolds
$\Sigma$.
These observables are associated with graphs in 2-dimensions, and
for the latter case there are additional observables that form a
Kac-Moody
algebra. The general solution of the classical equations of
 motion are also given.  Quantization is briefly discussed and
  some  quantum states are obtained  by Dirac quantization.
 The action of the observables on these states is given.

The action is constructed from
a zweibein $e_\alpha^i$, two Abelian gauge fields $A_\alpha^i$
($\alpha=1,2,3$, $i=1,2$), and $2N$ scalar fields $\pi_n$ and
$\phi_n$ $(n=1,2..,N)$:

\begin{equation}
 S=\int _M \bigl[\delta_{ij}e^i\wedge F^j +
 \epsilon_{ij}e^i\wedge e^j\wedge \pi_nd\phi_n \bigr]
 \end{equation}
where $F_{\alpha\beta}^i= \partial_{[\alpha}A_{\beta]}^i$.
The  model has $U(1)\times U(1)$ internal local gauge and 3d
 diffeomorphism invariance, (as well as global O(n) invariance).
   There is a degenerate metric on
 $M$, $g_{\alpha\beta}=e_\alpha^i e_\beta^j\delta_{ij}$ with the
 degeneracy direction given by the vector density
 $n^\alpha =
\epsilon^{\alpha\beta\gamma}e_\beta^ie_\gamma^j\epsilon_{ij}$.
(The zweibeins $e_\alpha^i$ however do not have their usual meaning
since
they do not rotate under the local internal group).
The dynamics (or rather lack thereof) in this theory is readily
understood
from the Hamiltonian perspective.

  With respect to a fixed foliation, the 2+1 form of this action is
\begin{equation} S=\int_R dt\int_{\Sigma}d^2x
\epsilon^{0ab}\bigl[-e_a^i\dot{A}^i_b
 + n^0\pi_n\dot{\phi}_n -A_0^i\partial_be_a^i
 + e_0^i(F_{ab}^i + 2\epsilon_{ij}e_a^j\pi_n\partial_b\phi_n)\bigl]
\end{equation}
where $a,b,..$ are world indices on $\Sigma$.
  The canonical phase space variables on the
 2d surfaces $\Sigma$ are then the pairs $(A_a^i,E^{ai})$ and
 $(\phi_n,\Pi_n)$, where $E^{ai} = \epsilon^{ab}e_b^i$ and
 $\Pi_n = n^0\pi_n(=\pi_n$det$E)$. The Hamiltonian action is
 \begin{equation}
 S=\int_{\Sigma\times R}dtd^2x \bigl[E^{ai}\dot{A}^i_a
 + \Pi_n\dot{\phi}_n + A_0^i\partial_aE^{ai}
 - N^a(E^{bi}F_{ab}^i + \Pi_n\partial_a\phi_n)\bigl],
 \end{equation}
where $A_0^i$ and $N^a\equiv e_0^ie^a_i$ appear as lagrange
mulitipliers.
 Varying with respect to them gives the constraints
\begin{equation}
 G^i\equiv\partial_aE^{ai}=0\ \ \ \ \ {\rm and} \ \ \ \ \ C_a\equiv
  E^{bi}F_{ab}^i
  +\Pi_n\partial_a\phi_n = 0
  \end{equation}
The Hamiltonian is a
 linear combination of these constraints
 \begin{equation}
 H=\int_\Sigma d^2x \bigl[N^a C_a + \Lambda^iG^i\bigr]
 \end{equation}
 where $N^a$ is the shift vector and $\Lambda^i(=-A_0^i)$ is the
gauge
 tranformation  function.
  The constraints (4) generate respectively  $U(1)\times U(1)$ gauge
transformations and  2d spatial diffeomorphisms of the phase space
variables. There are $n$  local degrees of freedom which are
associated
with the scalar fields and, depending on the spatial topology, there
are also
a finite number of topological degrees of freedom.
 Since there is no Hamiltonian constraint in the theory, there is no
 dynamics off the spatial surfaces $\Sigma$. Thus the action,
 though manifestly covariant in 3-dimensions, effectively
 gives a two dimensional field theory.

For $\Sigma$ with boundary the variation of the Hamiltonian (5)
gives rise to surface terms, and for Hamilton's equations
to be well defined one requires the vanishing on the boundary of
either the fluctuations of the
physical fields $(A,E,\phi,\Pi)$  {\it or} the gauge transformation
 functions $N^a, \Lambda^i$.
 Since we would like to discuss
physical observables associated with the boundaries, we make the
latter choice. (This is analogous to the fall  off conditions on the
lapse
and shift functions in general relativity which give rise to the
ADM observables for spatial slices with a boundary
\cite{adm,teitreg}).

 There is a natural set of gauge invariant physical
 observables of the model based on loops. If $\Sigma$ is without
 boundary, the configurations
 of the scalar fields $\phi_n(x,y)=$constants$=c_n$ define a set of
  $n$ loops
 $\gamma_n$ on $\Sigma$. The `internal' observables constructed using
 these loops are
\begin{equation}
 Q_n^i[A,\phi_1,...\phi_n](c_1,...c_n) =
 {\rm exp}[\int_{\gamma_n} dx^a A_a^i]
 \end{equation}
 \begin{equation}
  P_n^i[E,\phi_1,...\phi_n](c_1,...c_n) =
  \int_{\gamma_n} dx^a E^{bi}\epsilon_{ab}
\end{equation}
These satisfy the Poisson algebra
\begin{equation}
 \{Q_m^i,P_n^j\} = \int_{\gamma_m}ds\int_{\gamma_n}dt
\dot{\gamma}_m^a(s)\dot{\gamma}_n^b(t)\epsilon_{ab}
\delta^2(\gamma_m(s),\gamma_n(t))\delta^{ij}Q_m^i
 \end{equation}
 (no sum on $i,j$), and have vanishing Poisson brackets with
 the constraints (4). The invariance under the diffeomorphisms may be
 seen by computing the Poisson brackets, or by noting that the
 integrands are 1-densities made from phase space variables, which
are
 then integrated over paths also made from phase space variables.

 These observables have a number of interesting properties.
 They are diffeomorphism invariant versions of the observables in
electromagnetism \cite{loopem}.
 The $n$ loops determined by $\phi_n(x,y)=c_n$
  in general intersect to give a graph. One can
  for example, obtain graphs on $\Sigma$ where four lines intersect
  at a point by choosing the appropriate configurations of the
   scalar fields to make the loops.
 Increasing the number of scalar fields
in the action gives more loops on $\Sigma$, which in general
increases the number of vertices of the resulting graph.
Conversely, given for example {\it any} graph with $n$ vertices and
with four lines meeting at a vertex, one can find configurations of
$\phi_n$ which give that graph, and hence associate
to it observables of this theory via $Q_n^i$ and $P_n^i$.
(This can also be done for graphs where any number of lines meet at
a vertex by making an appropriate choice of loops via the
configuration of scalar fields).

When $\Sigma$ is a 2-manifold with a number of $S^1$ boundaries
there are  additional observables associated with each boundary
element.
  These  arise essentially due to the fall off conditions
  on $N^a, \Lambda^i$ discussed above. They are
 parametrized by arbitrary functions $\rho^i, \sigma^i$:
\begin{equation}
  p^i[E;\rho]=\int_\Sigma d^2x E^{ai}\partial_a\rho^i
  \end{equation}
\begin{equation}
q^i[A,E;\sigma]=\int_\Sigma d^2x
\bigl[\epsilon^{ab}A_a^i\partial_b\sigma^i
  - 2\phi_n\epsilon^i_{\ j} E^{bj}
  \partial_b({\sigma^i \Pi_n\over {\rm det}E})\bigr]
  \end{equation}
 (no sum on $i$ in (9-10)). The Poisson brackets of these  with
 the constraints give  volume and boundary terms. The former are
  proportional to the constraints, while the
 latter  vanish (due to the vanishing of $\Lambda^i, N^a$ on
 the boundary).  The Poisson bracket of (10) with the
 diffeomorphism  constraint is proportional to the coefficient of
 $e^i_0$ in (2)(which is proportional to the diffeomorphism
constraint).

  There are still the interior observables (6-7), again specified
by $\phi_n = c_n$,  but these latter conditions can now also give
open
curves $\gamma[a,b]$ that begin and end at boundary points $a,b$,
which may
 lie on the same or different boundary components.
These are still invariant, again because the parameters of
the symmetry transformations $N^a,\Lambda^i$
vanish on the boundaries.

 The Poisson algebra of the $q^i,p^i$ with themselves, and with the
observables (6-7) based on curves $\gamma[a,b]$ is
 \begin{equation}
 \{q^i[A,E;\sigma],p^j[E;\rho]\} =
 \delta^{ij}\int_{\partial\Sigma} d\theta \sigma^i\partial_\theta
\rho^i
 \end{equation}
 \begin{equation}
 \{Q^i_n[A,\phi_n](c_n),p^j[E;\rho]\} = \delta^{ij}
 (\rho^i (b)-\rho^i (a))Q^i_n[A,\phi_n](c_n)
 \end{equation}
 \begin{equation}
 \{P^i_n[E,\phi_n](c_n),q^j[A,E;\sigma]\}=\delta^{ij}
 (\sigma^i (b)-\sigma^i (a)).
 \end{equation}
If the functions $\rho^i, \sigma^i$ are chosen such that
they reduce to $exp{iN\theta}/\sqrt{2\pi}$ on a boundary component
$S^1$,
(where $\theta$ is the
coordinate on $S^1$), then the observables (9-10)
may be labelled by integers $M,N$ associated with the $\rho,\sigma$
boundary values.
 The Poisson brackets (11) may then be rewritten as
\begin{equation}
\{q^i_M,p^j_N\}= iM\delta_{M+N,0}\delta^{ij},
\end{equation}
where the boundary values of $\rho^i,\sigma^i$ have been substituted
into the rhs of (11).
Equations (14) are two copies of a U(1) Kac-Moody algebra associated
with every $S^1$ element of the boundary $\partial\Sigma$.
(The remaining Poisson brackets $\{Q_n^i[A,\phi_n](c_n),$
$q^j[E,A;\sigma]\}$ and $\{q^i[E,A;\sigma],q^j[E,A;\rho]\}$
 give volume and boundary terms. The former are proportional to
 the constraints and the latter are zero
when the scalar field phase space variables vanish on the
boundaries).

Without the scalar fields the theory becomes topological and the
loop observables (6-7) become trivial. In this case only the
Kac-Moody observables (9-10) and any topological observables
associated with non-contractible loops remain.

Although an infinite number of constants of the motion are given
above, the question of integrability  requires further study.
 A hint that the theory may be integrable comes from
looking at the Hamilton-Jacobi equations. Just as in
unconstrained mechanics, the solution of the HJ
equations provide a way of mapping the trajectories to the initial
data. In the present case, since there is no dynamics, the
trajectories are gauge orbits, and a solution of the HJ equations
with the appropriate number
of integration momenta can provide a map
to the unconstrained gauge invariant variables. A procedure to do
this
has recently been given  by Newman and Rovelli \cite{tc} and we now
 apply it here.
The HJ equation for the Gauss constraint is
\begin{equation}
\partial_a {\delta S[A^i]\over \delta A_a^i}=0,
\end{equation}
which has solution $S[A^i;u^i]=
\int_\Sigma \epsilon^{ab}A_a^i\partial_b u^i$,
 where the two functions $u^i$ are the `reduced integration momenta'
on the solutions of the Gauss constraint.
{}From this the new `coordinates' conjugate to $u^i$ are
$q_{u^i} = \delta S/\delta u^i$ = $\epsilon^{ab}F_{ab}^i$,
and the old momenta are $E^{ai} = \delta S/\delta A_a^i =
\epsilon^{ab}\partial_b u^i$. The Gauss law has now been eliminated
 and substituting these into $C_a$ (4) gives the
 `reduced' diffeomorphism constraint
 \begin{equation}
 C_a = q_{u^i}\partial_a u^i + \Pi_n\partial_a\phi_n.
 \end{equation}
  The HJ equation corresponding to this is
\begin{equation}
(\partial_a u^i){\delta S[u^i,\phi_n]\over \delta u^i} +
(\partial_a\phi_n) {\delta S[u^i,\phi_n]\over \delta \phi_n}=0
\end{equation}
The solution of (17) should have the appropriate number of
integration momenta, namely $n$, since there are $n+2$ functional
derivatives in it and two constraints $C_a$ to be solved.
The solution is
 \begin{equation}
 S[u^i,\phi_n; P_n] =
\int_\Sigma d^2x \epsilon^{ab}\epsilon_{ij}
(\partial_a u^i)(\partial_b u^j)\phi_n P_n(u(x))
\end{equation}
 where $P_n$ are the integration momenta.
 That this is a solution may be verified by an explicit calculation,
 or the observation  that the integrand is a density made from the
 phase space coordinates.
  It  may be rewritten as
$S[u^i,\phi_n; P^n]=\int_\Sigma d^2u \phi_n(x(u))P_n(u)$
in terms of coordinates on $\Sigma$ defined by the two functions
$u^i$. These forms of $S$ give the canonical transformation to
the new reduced phase space variables $Q_n,P_n$ in terms of the old:
 \begin{eqnarray}
Q_n(x)&=& \phi_n(u^{-1}(x)), \nonumber \\
P_n(x)&=& \Delta^{-1} \Pi_n(u^{-1}(x)) \nonumber \\
q_{u^i}(x)&=& -\epsilon^{ab}\epsilon_{ij}
(\partial_a\phi_n)(\partial_bu^j)P_n(u(x))
 \end{eqnarray}
where $\Delta = \epsilon^{ab}\epsilon_{ij}
 (\partial_a u^i)(\partial_b u^j)$.
This gives the complete solution of the canonical equations of
 the motion! Thus this model is integrable.
 By inverting
 (19) the invariant observables $Q_n,P_n$ may be rewritten
 as functionals of the original phase space variables.

The solution (19) may be given an interesting geometrical
interpretation. The field lines of $E^{ai}$ are tangent to
the loops (or loops and open curves for $\Sigma$ noncompact)
determined by $u^i$=constants \cite{tc}. The solution consists in
simply
evaluating the scalar field variables on the electric field lines.
  This is in contrast to the other invariant observables that we
discussed where the gauge fields are evaluated on loops determined
by the scalar fields. As a final remark, an alternative way to
proceed
 is to solve the reduced diffeomorphism constraint (16) after
imposing the
 coordinate fixing conditions $u^i=x^i$. This gives the constants of
the
 motion $(\phi_n(u),\Pi_n(u))$ with
 $q_{u^i}(u) = -\Pi_n(u)\partial_i\phi_n(u)$ as the general solution.
This is equivalent to (19) in these coordinates.

 The Dirac quantization of the model may be carried out by
converting the constraints into operator conditions on
wave functions. Since the constraints are linear
in the momenta there are no  operator ordering ambiguities.
In the configuration representation  these conditions are
 \begin{eqnarray}
&& \partial_a{\delta \over  \delta A_a^i}\Psi[A^i,\phi] =0 \\
&& \bigl(F_{ab}^i{\delta \over  \delta A_b^i}   +
 \partial_a \phi_n {\delta \over \delta\phi_n}\bigr)\Psi[A^i,\phi] =
0
\end{eqnarray}

These equations are solved by the functional
\begin{equation}
\Psi [A^i,\phi_n](c_1,..c_n)= {\rm exp} [\int_{\gamma_n} A_a^i dx^a
].
\end{equation}
 This is the line integral around the graph $\gamma_n$ determined
 by the collection of intersecting loops $\phi_n = c_n$ and is
identical
  to the observable (6).
Among the  special cases of (22) are quantum states of the model
  associated with any $n$ vertex graph where four lines meet at a
  vertex. These may be associated with Feynman graphs of $\lambda
\phi^4$
  field theory \cite{gins}. Such graphs can always be constructed
  from appropriate  configurations of the scalar fields that give a
  collection of loops such that every pair has two intersections.

 The operators corresponding
 to the observables
$Q_n^i$ and $P_n^i$ have a well defined action on these states.
In particular $P_n^i$ is diagonal,
\begin{eqnarray}
\hat{P}_m^i \Psi[A^j,\phi] &=& \int_{\gamma_m} dx^a \epsilon_{ab}
\frac{\delta}{\delta A_b^i}{\rm exp}[\int_{\gamma_n} A_a^j dx^a
]\nonumber
\\  &=& \int_{\gamma_m}ds\int_{\gamma_n}dt
 \dot{\gamma}_m^a(s)\dot{\gamma}_n^a(t)\epsilon_{ab}
 \delta^2(\gamma_m(s),\gamma_n(t))\delta^{ij} \Psi[A,\phi],
 \end{eqnarray}
and the eigenvalue is the number of intersections of the graph
$\gamma_m$
associated with $\hat{P}_m^i$ with the graph $\gamma_n$ associated
with
 $\Psi$.  The action of $\hat{Q}_n^i$ is by multiplication
 and it acts like a raising operator:
\begin{equation}
\hat{Q}_m^i\Psi[A^i,\phi_n] = {\rm
exp}[\int_{\gamma_m+\gamma_n}A_a^idx^a]
\end{equation}
 The Kac-Moody observable (9) annihilates the states (22)
 that are based on loops. This is
expected because these observables are the `integrated by parts
versions' of the constraints. But for the case of $\Sigma$ with
boundary,
there are also physical states (22) based on curves $\gamma[a,b]$
that begin and end on a boundary component. The action of the
Kac-Moody
observable (9) on these is
\begin{equation}
 \hat{p}^i[E,\rho]\Psi=(\rho(b)-\rho(a))\Psi
\end{equation}
The action of the second observable (10) is more complicated since it
involves det$E$, and a careful calculation is required.

The main result in this paper is the observation that it is
possible to obtain a 2d field theory from a 3d non-topological theory
in
which there is no dynamics of the local degrees of freedom. The
2d theory obtained is described by the constraints
(4) on the phase space of scalar fields and $U(1)\times U(1)$
Yang-Mills fields. This group was chosen to get a 3-dimensional
action with a zweibein. But the 2-metric of this theory
 constructed from the phase space zweibeins $E^{ai}$ does
 not give the conformal factor: the metric is not conformally flat
 but just flat. This is due to the fact that there is no field
 corresponding to the 2d spin connection, and so the theory
is not one of 2d gravity (and is not conformally invariant).
Nevertheless, there are an infinite number of  conserved
charges for the theory and it is integrable. Furthermore, the theory
also
has a close relationship with the kinematical aspects of one Killing
field
reductions of general relativity \cite{vhj} in the Ashtekar
formalism.
  It would be of interest to see if there are
 related 3d theories that give other 2d theories
in a similar way.

I would like to thank Andrei Barvinsky and Nemanja Kaloper  for
helpful
comments on the manuscript. This work was supported by NSERC of
Canada.

\end{document}